\documentclass[twocolumn,pra,twocolumn,superscriptaddress]{revtex4-2}
\usepackage{color}
\usepackage{amsmath,amsthm,mathrsfs,dsfont}
\usepackage{graphicx}
\usepackage{tikz}
\usepackage{xr}
\usepackage{xc}
\usepackage{float}
\usepackage{bbm}
\usepackage{epsfig}
\usepackage{verbatim}
\usepackage{array}
\usepackage{setspace}
\usepackage{bbding}
\usepackage{amssymb}% http://ctan.org/pkg/amssymb
\usepackage{pifont}% http://ctan.org/pkg/pifont
\usepackage{braket}
\usepackage{tikz}
\usepackage{qcircuit}

\usepackage{newtxtext,newtxmath}

\usepackage[unicode=true,
 bookmarks=true,bookmarksnumbered=false,bookmarksopen=false,
 breaklinks=false,pdfborder={0 0 1},backref=false,colorlinks=true]
 {hyperref}
\hypersetup{
 linkcolor=magenta, urlcolor=blue, citecolor=blue, pdfstartview={FitH}, unicode=true}
 
\usepackage{cleveref} 
\usepackage{color}
\crefname{section}{Sec.}{Secs.}
\crefname{figure}{Fig.}{Figs.}
\Crefname{figure}{Figure}{Figures}

\crefname{table}{Table}{Tables}
\crefformat{table}{Table~#2#1#3}
\crefformat{table}{Table~#2#1#3}
\crefname{equation}{Eq.}{Eqs.}

\crefname{algorithm}{Algorithm}{Algorithms}
\crefformat{algorithm}{Algorithm~#2#1#3}
\crefformat{algorithm}{Algorithm~#2#1#3}

\makeatletter
\@ifundefined{textcolor}{}
{%
 \definecolor{BLACK}{gray}{0}
 \definecolor{WHITE}{gray}{1}
 \definecolor{RED}{rgb}{1,0,0}
 \definecolor{GREEN}{rgb}{0,1,0}
 \definecolor{BLUE}{rgb}{0,0,1}
 \definecolor{CYAN}{cmyk}{1,0,0,0}
 \definecolor{MAGENTA}{cmyk}{0,1,0,0}
 \definecolor{YELLOW}{cmyk}{0,0,1,0}
}

\usepackage{amsfonts}\usepackage{dcolumn}\usepackage{bm}\usepackage{epstopdf}
\usepackage{algorithm}
\usepackage{algpseudocode}
\usepackage{booktabs}

\hypersetup{urlcolor=blue}

\makeatother

\newcolumntype{C}[1]{>{\centering\arraybackslash$}p{#1}<{$}}

\graphicspath{{fig/}}

\begin{document}

\widetext
\title{Clustering-Based Sub-QUBO Extraction for Hybrid QUBO Solvers}% Force line breaks with \\

\author{Wending Zhao}%
\affiliation{Huayi Boao (Beijing) Quantum Technology Co., Ltd., 100176, Beijing, People's Republic of China }

\author{Gaoxiang Tang}
\email{tgx24@mails.tsinghua.edu.cn}
\affiliation{Center for Quantum Information, IIIS, Tsinghua University, Beijing 100084, People's Republic of China}

\begin{abstract} 
Quantum Approximate Optimization Algorithm (QAOA) offers a promising approach for solving quadratic unconstrained binary optimization (QUBO) problems. However, the size of the solvable problem is limited by the number of qubits. To leverage noisy intermediate-scale quantum (NISQ) devices to solve large-scale QUBO problems, one possible way is to decompose the full problem into multiple sub-problems, which we refer to as the Sub-QUBO Formalism.
In this work, we enhance this formalism by proposing a sub-QUBO extraction protocol. To do so, we define a measure to quantify correlations between variables and use it to build a correlation matrix. This matrix serves as the input for clustering algorithms to group variables. Variables belonging to the same group form sub-QUBOs and are subsequently solved using QAOA.
Our numerical analysis on several classes of randomly generated QUBO problems demonstrates that this grouping method outperforms previous approaches in terms of objective function values, while maintaining a comparable number of quantum subroutine calls. This method offers wide applicability for solving QUBO problems on NISQ devices.

\end{abstract}

\maketitle

%\tableofcontents

\section{Introduction}
Quadratic unconstrained binary optimization (QUBO) is a fundamental \textit{NP}-hard problem in combinatorial optimization \cite{kochenberger2014unconstrained,riandari2021quantum}. Various classical heuristic algorithms can approximate the solutions of QUBO problems. These algorithms have been extensively studied, particularly those based on local search methods, such as simulated annealing \cite{alkhamis1998simulated,katayama2001performance} and tabu search \cite{glover1998adaptive,glover1998tabu}.

In the quantum domain, solving QUBO is equivalent to finding the ground state of an Ising Hamiltonian. Several quantum algorithms have been developed for this purpose. Quantum Signal Processing (QSP) filter algorithms \cite{lin2020near} and Lindbladian-based approaches \cite{chen2023quantum, ding2024single} require fault-tolerant quantum computers to find low-energy states. Algorithms based on the Hadamard test and classical Fourier transform are applicable to early fault-tolerant quantum computers \cite{lin2022heisenberg}, while quantum annealing requires specialized hardware \cite{king2025beyond}. Variational quantum algorithms, such as the Quantum Approximate Optimization Algorithm (QAOA), can be implemented on near-term intermediate-scale quantum (NISQ) devices \cite{farhi_quantum_2014,blekos2024review,bharti2022noisy}. QAOA encodes the QUBO problem into a cost Hamiltonian and applies alternating layers of time evolution governed by this Hamiltonian and a transverse field. By optimizing $O(p)$ parameters, where $p$ denotes the number of layers, the algorithm aims to approximate the ground state solution \cite{farhi_quantum_2014,wecker2016training}.

However, the practical deployment of QAOA on current NISQ devices is constrained by the limited number of high-fidelity qubits, shallow circuit depths, and noise accumulation, which restrict its applicability to small-scale problems \cite{bharti2022noisy, Pelofske2024, Omanakuttan2025}. 
To overcome these limitations, researchers have applied the concept of problem decomposition to QAOA, resulting in the development of hybrid QUBO solvers with sub-QUBO extraction \cite{rosenberg2016building,booth2017partitioning, shaydulin2019hybrid, warren2020solving, liu2022leveraging, yatabe2024partitioning}. 

Within the sub-QUBO formalism, the main QUBO objective is iteratively decomposed into smaller sub-QUBOs by grouping variables. Each sub-QUBO problem is solved using a quantum subroutine, and their solutions are combined and refined to a local minimum via classical optimization. 
A critical component of the sub-QUBO formalism is the grouping strategy, as the quality of grouping directly impacts both the quality of the final solution and the convergence rate of the algorithm.
Various strategies have been proposed for variable partitioning, including random selection \cite{rosenberg2016building}, sorting variables based on the impact of each variable on the objective function (termed Impact-Indexing)\cite{booth2017partitioning}, and extracting the most undetermined variables from a pool of previous quasi-solutions (termed Certainty-Degree)\cite{rosenberg2016building,atobe_hybrid_2022}.

In this study, we propose a clustering-based variable grouping scheme in sub-QUBO formalism. Firstly, we construct a correlation matrix to capture the joint contributions of variables to the objective function. The correlation matrix can be derived from the QUBO matrix and the current solution. Then, we employ the multi-view spectral clustering method \cite{von2007tutorial,NIPS2011_31839b03} to group variables into clusters according to their correlations. 
We evaluate the performance of our algorithm on Max-Cut problems of randomly generated regular graphs and Erdős–Rényi graphs. Our numerical simulations demonstrate that our method consistently achieves lower objective function values than existing approaches.

The remainder of this article is organized as follows: In \cref{sec:framework}, we review the concept of QAOA for QUBO problems and the sub-QUBO formalism. In \cref{sec:cluster-grouping}, we introduce clustering as a grouping method in the sub-QUBO formalism. In \cref{sec:numerics}, we present numerical simulations on datasets of Max-Cut problems, and demonstrate the performance of different algorithms. Finally, in \cref{sec:conclusion}, we summarize the key findings of this work.

\begin{figure*}
    \centering
    \includegraphics[width=1\linewidth]{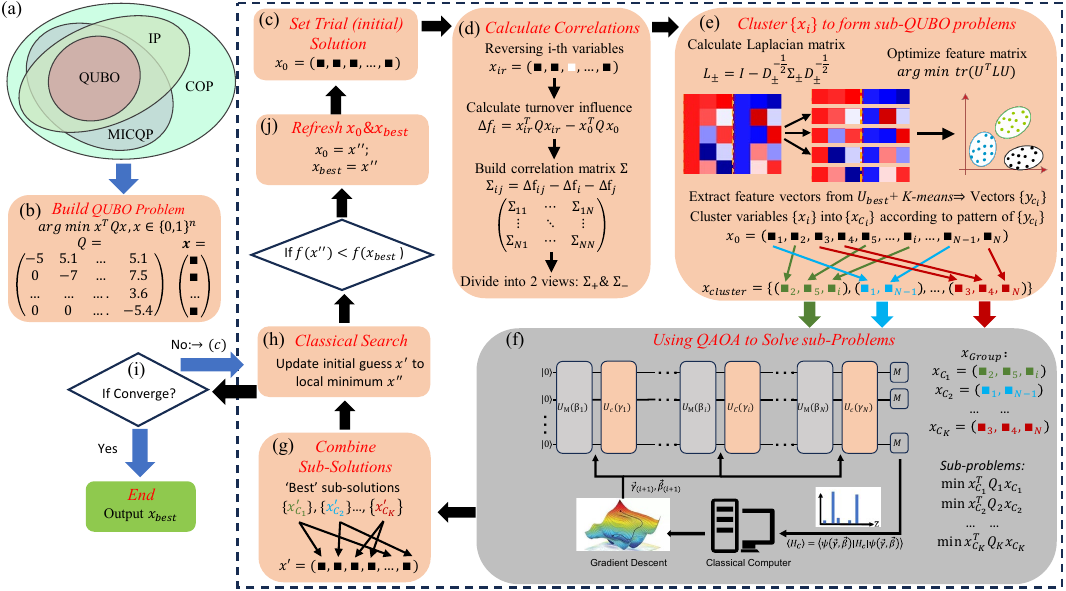}
    \caption{The pipeline of utilizing the sub-QUBO formalism to address combinatorial optimization problems (COPs). (a) A Venn diagram of QUBO and other COPs. Many COPs can be formulated as QUBOs. (b) The initial step in solving these problems is the construction of a QUBO representation. (c) An initial guess, denoted as $x_0$, serves as the starting point for the iterative algorithm. (d) The correlation matrix $\Sigma$ is computed based on current solution $x_0$ and QUBO matrix $Q$. This matrix is further segmented into a positive component, $\Sigma_{+}$, and a negative component, $\Sigma_{-}$. (e) The grouping of independent variables is achieved through the construction of the Laplacian matrix, $L_{\pm}$, followed by clustering its eigenvectors using the K-means algorithm. (f) After clustering, the original problem is decomposed into sub-QUBO problems, represented as $\min x_{C_{k}}^{T} Q_{k} x_{C_{k}}$. Each sub-problem is then solved using the QAOA algorithm on quantum hardware. (g) The optimal solutions of each sub-problem, denoted as $x_{C_{k}}^{\prime}$, are integrated to update the current global solution, represented as $x^{\prime}$. (h) A classical local search performed with initial guess $x^\prime$ results in a local minimum of $x^{\prime\prime}$. (i) If the convergence condition is met, the iterative loop (c-h) terminates, resulting in the final output.  (j) If $x^{\prime\prime}$ outperforms the previous best $x_{\text{best}}$, $x_{\text{best}}$ is updated and used as the initial guess for a new iteration. }
    \label{fig:hybrid_qubo_flow}
\end{figure*}

\section{Hybrid Quantum Algorithms For Solving QUBO Problems}
\label{sec:framework}
\subsection{QUBO Problems and QAOA}
The QUBO model is a versatile mathematical framework for representing optimization problems where the objective function is quadratic, and the decision variables are binary (\(x_i \in \{0, 1\}\)). The general form of a QUBO problem is expressed as:

\begin{equation}
    \min \mathbf{x}^T Q \mathbf{x} + \mathbf{c}^T \mathbf{x},
\end{equation}
where \(\mathbf{x}\) is a vector of binary variables, \(Q\) is a square matrix representing the coefficients of the quadratic terms, and \(\mathbf{c}\) is a vector containing the coefficients of the linear terms. Given the binary nature of the variables (\(x_i \in \{0, 1\}\)), the linear terms can be equivalently transformed into quadratic terms, enabling the entire problem to be reformulated in a purely quadratic form. The significance of the QUBO framework lies in its ability to encompass a broad range of optimization problems. Many COPs of practical significance can be reformulated by encoding constraints and objectives into binary variables and a quadratic cost matrix \cite{glover2018tutorial}.

For example, Mixed Integer Constrained Quadratic Programming (MICQP) problems can be transformed into QUBO by mapping real-valued variables into finite binary representations and adding penalty terms to manage constraints \cite{buonaiuto2023best,richoux2023learning}. 
Therefore, the QUBO framework, as a fundamental framework for solving COPs, has received widespread attention and application.

To utilize QAOA for solving the QUBO problem, the objective function is firstly transformed into an Ising Hamiltonian \cite{lucas2014ising} represented as:
\begin{equation}
H_C(\mathbf{s}) = \sum_{i} h_i s_i + \sum_{i<j} J_{ij} s_i s_j,
\label{ising_ham}
\end{equation}
where \(s_i\) are spin variables that can take values \(\pm 1\), \(h_i\) are external magnetic fields, and \(J_{ij}\) are interaction coefficients between spins \(i\) and \(j\). The mapping of binary variables \( x_i \in \{0, 1\} \) of the QUBO model to spin variables \( s_i \in \{-1, +1\} \) of the Ising model is $x_i = \frac{1 + s_i}{2}$. 

Solving the ground state energy of the Ising model is a \textit{NP}-hard problem \cite{kochenberger2014unconstrained,riandari2021quantum}. Classical heuristic algorithms can provide approximate solutions \cite{alkhamis1998simulated,katayama2001performance,glover1998adaptive,glover1998tabu}. Specialized computers, such as coherent Ising machines, offer alternative methods that can yield accurate approximations at high speeds \cite{haribara2022quantum, king2025beyond}. Additionally, quantum computing provides another promising approach for tackling Ising problems, as demonstrated in recent experiments using both analog \cite{king2016observation,mohseni2023isingtype,georgescu2021quantum,johnson2012quantum,puri2022quantum} and digital quantum simulations \cite{kim2011quantum,barends2015digital,mi2023quantum,satzinger2023realizing}.

QAOA \cite{farhi_quantum_2014, blekos2024review} is a variational quantum algorithm that approximates the ground state of a Hamiltonian using parameterized quantum circuits. 
As illustrated in \cref{fig:hybrid_qubo_flow}(f), the QAOA involves two key components: the cost Hamiltonian \(H_C\), which corresponds to the Ising Hamiltonian in our case, and the mixer Hamiltonian \(H_M = \sum_i \sigma_x^i\) that facilitates exploration of the solution space.  

The QAOA algorithm begins with the preparation of an initial state $\ket{\psi_0} = H^{\otimes n} \ket{0}^{\otimes n}$, which is an equal superposition of all possible states. The algorithm then alternates the application of time evolution operators corresponding to $H_M$ and $H_C$, parameterized by $2p$ parameters $\beta_i$ and $\gamma_i$ ($i=1,\ldots,p$), to this initial state, resulting in the following form for the variational state:
\begin{equation}
\ket{\psi(\vec{\gamma}, \vec{\beta})} = e^{-i \beta_p H_M} e^{-i \gamma_p H_C} \cdots e^{-i \beta_1 H_M} e^{-i \gamma_1 H_C} \ket{\psi_0}.
\end{equation}

Its measurement yields the expected value of the cost Hamiltonian, which is expressed as:
\begin{equation}
\langle H_C \rangle = \bra{\psi(\vec{\gamma}, \vec{\beta})} H_C \ket{\psi(\vec{\gamma}, \vec{\beta})}.
\end{equation}
Classical optimization techniques are then used to find the parameters \( \vec{\gamma}^* \) and \( \vec{\beta}^* \) that minimize \( \langle H_C \rangle \). Measurement of $\ket{\psi(\vec{\gamma}^*, \vec{\beta}^*)}$ on the computational basis results in the approximate ground state of $H_C$ with high probability.

Despite growing interest, the experimental implementation of QAOA remains constrained by hardware limitations. Recent analog realizations using neutral-atom arrays have scaled to hundreds of qubits and demonstrated promising results on combinatorial problems \cite{Ebadi2022}, while gate-based implementations on superconducting and trapped-ion devices are typically limited to fewer than 100 qubits and shallow circuit depths ($p \leq 3$) \cite{Pelofske2024}. These constraints significantly limit the problem sizes that can be realistically addressed. As a result, QAOA has yet to demonstrate consistent performance advantages over classical heuristics in practical settings \cite{DePalma2023}. These limitations motivate hybrid frameworks that leverage both classical solvers and quantum subroutines to extend QAOA’s applicability.

\subsection{The Sub-QUBO Formalism}
To address the scalability challenges of quantum algorithms in large QUBO problems, the sub-QUBO formalism has been introduced \cite{rosenberg2016building,booth2017partitioning}. As shown in \cref{fig:hybrid_qubo_flow}, this approach iteratively tackles the problem by decomposing it into smaller sub-QUBOs, each of which is solved with a quantum subroutine. The results of these sub-problems are then combined and refined to generate a new candidate solution for the subsequent iteration. 

The algorithm begins by initializing a solution vector derived from a classical search. Each iteration starts by partitioning the binary variables into distinct groups. For each group, the selected variables are mapped onto a sub-QUBO problem by re-expressing the objective function in terms of these variables while fixing the variables outside the group. Let $S$ denote the set of selected variables, the sub-QUBO objective function is expressed as:
\begin{equation}
    f_S(x_S) 
    = \sum_{i \in S} (Q_{ii} + d_i) x_i + \sum_{\substack{i, j \in S \\ i < j}} Q_{ij} x_i x_j,
\end{equation}
where $d_i = \sum_{j \notin S} (Q_{ij} + Q_{ji}) x_j^*$. Here, $x_i$ represents the variables in $S$, while $x_j^*$ refers to a set of fixed parameters outside $S$.

Each sub-QUBO is solved using a quantum subroutine, and the solution updates the current candidate solution. However, because of the approximate nature of the quantum subroutine, the updated solution may not be a local minimum. Consequently, a classical search is performed to force the solution to a local minimum. If the QAOA solution exceeds the previously best solution, the algorithm updates the optimal solution and proceeds to the next iteration. If no improvement occurs over multiple iterations, the convergence criterion is met, terminating the process.

In this formalism, three components allow flexible implementation choices. The first is the grouping method, discussed in detail in \cref{subsec:group-sub-QUBO}. The second component is the sub-QUBO solver, which can be quantum annealing, QAOA, or classical algorithms. In this work, we focus primarily on QAOA. Finally, the classical search that forces the solution to a local minimum can also switch between multiple choices. In our work, we employ a simple greedy local search to emphasize the role of the quantum subroutine, while previous studies have explored other classical search methods, including the tabu search \cite{glover1998tabu,wang2012path} as used in Ref.~\cite{booth2017partitioning}. A discussion on the performance of tabu search within this formalism is provided in \cref{app:tabu}.

\subsection{Grouping Methods in the Sub-QUBO Formalism}
\label{subsec:group-sub-QUBO}

The effectiveness of hybrid annealing algorithms critically depends on how sub-QUBOs are extracted. Ref.~\cite{rosenberg2016building} introduced the hybrid annealing framework and several grouping strategies, including random selection and an Impact-Indexing approach (originally referred to as gains-based). The Impact-Indexing method measures the 'impact' of each variable by evaluating how much the objective function changes when that variable is flipped. Initially, Ref.~\cite{rosenberg2016building} formed sub-QUBOs by choosing the variables with the largest negative impacts. 

Subsequently, Ref.~\cite{booth2017partitioning} refined the Impact Indexing method by sorting all variables according to their computed impacts and then grouping contiguous variables so that variables within each group have approximately the same impact. Treating each of these groups as a separate sub-QUBO mitigates the local optima issues encountered in the original approach.

Another strategy, proposed in Ref.~\cite{atobe_hybrid_2022}, introduces the concept of certainty degree. Here, the algorithm keeps a pool of \( N_S \) local minima. For each variable \(i\), its degree of certainty \( d_i \) is defined as
\begin{equation}
    d_i = \left| \frac{N_S}{2} - c_i \right|,
\end{equation}
where 
\(
c_i = \sum_{k=1}^{N_S} x_i^{(k)},
\)
and \( x_i^{(k)} \) is the \(i\)-th component of the \(k\)-th solution vector in the pool. This method can be interpreted as a specialized form of diversity-driven tabu search~\cite{glover_diversification-driven_2010}, in which sub-QUBO solutions serve as perturbation mechanisms.

Both the refined Impact-Indexing and Certainty-Degree approaches share a similar rationale: variables of comparable impact or certainty are grouped together and solved within the same sub-QUBO. While these strategies have been demonstrated to be effective across various datasets, an even more intuitive method involves grouping strongly correlated variables. In such an approach, updating variables within one sub-QUBO is expected to minimally affect variables outside that group. As described in \cref{sec:cluster-grouping}, this idea can be realized using a clustering-based grouping method, which systematically forms sub-QUBOs by leveraging correlation measures among variables.

\section{Clustering-Based Grouping Method}
\label{sec:cluster-grouping}

\subsection{Correlation Matrix}
\label{subsec:correlation_mat}

To characterize the interactions between variables and enable efficient grouping in optimization, we define the correlation matrix based on the QUBO matrix \( Q \) and a given solution \( \mathbf{x} \). 

The initial step involves calculating the incremental change in the objective function, \( f \), when a single variable \( i \) is flipped. This change, denoted \( \Delta f_i(\mathbf{x}; Q) \), is expressed as:
\begin{equation}
 \begin{split}
    \Delta f_i(\mathbf{x}; Q) = & f(x_1, x_2, \ldots, x_i \oplus 1, \ldots, x_{N}; Q) \\
     & - f(\mathbf{x}; Q).
      \end{split}
\end{equation}

Similarly, \( \Delta f_{ij}(\mathbf{x}; Q) \) represents the change in the objective function when both variables \( i \) and \( j \) are flipped simultaneously, as defined by:
\begin{equation}
    \begin{split}
        \Delta f_{ij}(\mathbf{x}; Q) = & f(x_1, \ldots, x_i \oplus 1, \ldots, x_j \oplus 1, \ldots, x_{N}; Q) \\
        & - f(\mathbf{x}; Q).
    \end{split}
\end{equation}

The correlation between variables \( i \) and \( j \) is then defined by:
\begin{equation}
\label{equation:correlation}
 \begin{split}
    \Sigma_{ij}(\mathbf{x}; Q) & = \Delta f_{ij}(\mathbf{x}; Q) - \Delta f_i(\mathbf{x}; Q) - \Delta f_j(\mathbf{x}; Q) \\
    & = (-1)^{x_i + x_j} Q_{ij}.
    \end{split}
\end{equation}

To form clusters, variables with minimal absolute correlation are separated into different groups, thus limiting inter-group dependencies. By segmenting weakly correlated variables, each sub-QUBO can be optimized more independently, reducing the computational overhead associated with large-scale QUBO problems.
In this formulation, the element \(\Sigma_{ij}(\mathbf{x}; Q)\) of the correlation matrix captures how flipping both variables \(i\) and \(j\) simultaneously affects the objective function relative to flipping them individually. A positive \(\Sigma_{ij}(\mathbf{x}; Q)\) implies that flipping \(i\) and \(j\) together increases the objective more than the sum of their individual flips—an effect termed a “repulsive” interaction, since it is penalized in a minimization context. Conversely, a negative \(\Sigma_{ij}(\mathbf{x}; Q)\) corresponds to an “attractive” interaction, where flipping both variables either leads to a smaller increase in the objective or even a net decrease. 

\subsection{Clustering}
\label{subsec:cluster}
To form clusters, variables with minimal absolute correlation are separated into different groups, thus limiting inter-group dependencies. By segmenting weakly correlated variables, each sub-QUBO can be optimized more independently, reducing the computational overhead associated with large-scale QUBO problems.

Several algorithms are feasible for clustering a correlation matrix, including hierarchical clustering\cite{murtagh2012algorithms}, density-based spatial clustering of applications with noise (DBSCAN)\cite{ester1996density}, spectral clustering\cite{von2007tutorial}, and the Signed Laplacian method \cite{knyazev2017signed, doi:10.1137/1.9781611972801.49}. 
However, the presence of both positive and negative correlations in our case, along with the objective of minimizing both, complicates the application of these standard methods.

\begin{figure*}
    \centering
    \includegraphics[width=1.0\linewidth]{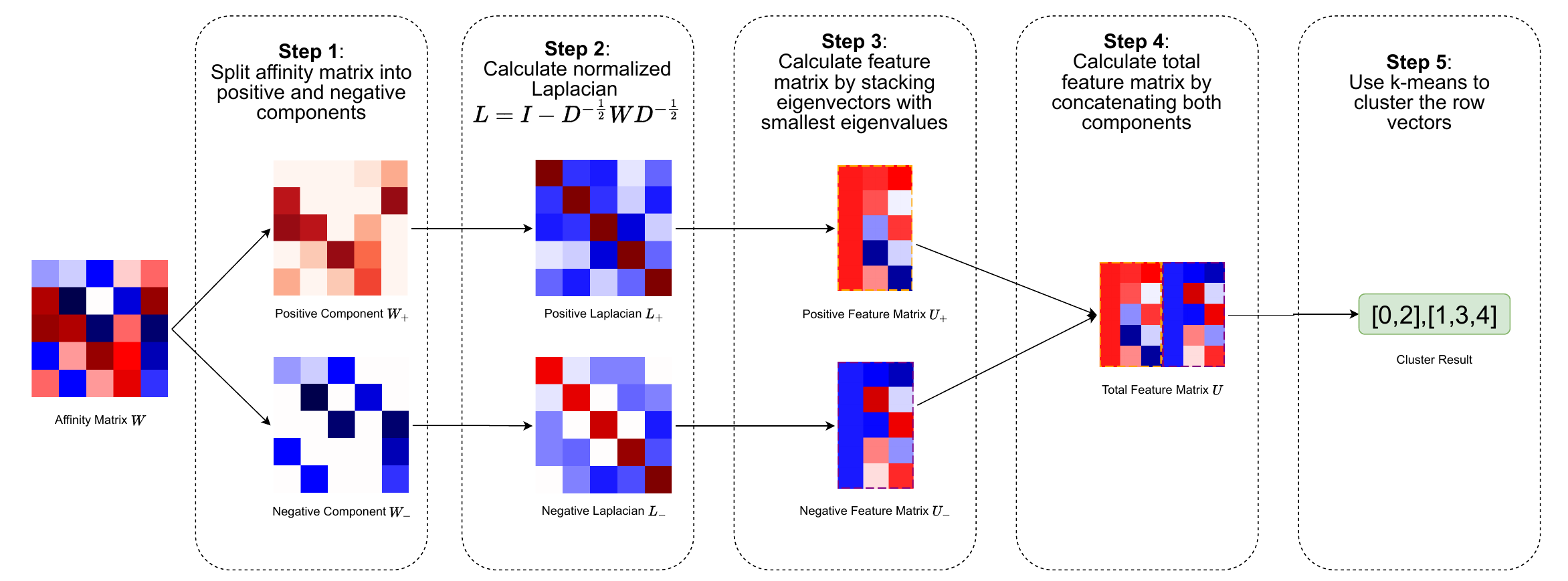}
    \caption{Diagram of feature-concatenation multi-view cluster}
    \label{fig:f-c_diagram}
\end{figure*}

To address this, we adopt a multi-view framework for clustering \cite{bickel2004multi}, as illustrated in \cref{fig:f-c_diagram}. In this approach, the correlation matrix \(\Sigma\) is decomposed into a positive view \(\Sigma_{+}\) and a negative view \(\Sigma_{-}\), corresponding to repulsive and attractive forces, respectively. Feature matrices are then extracted from each view through spectral embedding. Concretely, the normalized Laplacian for each view is defined as  
\begin{equation}
    L_{\pm} = I - D_{\pm}^{-\tfrac{1}{2}}\Sigma_{\pm}D_{\pm}^{-\tfrac{1}{2}},
\end{equation}
where \(D_{\pm}\) is a diagonal matrix whose entries are the degrees of the nodes for the respective view.

Each \(L_{\pm}\) is positive semidefinite, and its eigenvalues and eigenvectors play a key role in spectral clustering \cite{von2007tutorial}. In particular, the smallest non-zero eigenvalue, known as the Fiedler value, and its associated eigenvector (the Fiedler vector) provide insights into the clustering structure by grouping nodes with similar vector values. To extract the \(m\) most informative features for clustering, we select the first \(m\) eigenvectors \(u_1, u_2, \ldots, u_m\) of \(L_{\pm}\) and place them as columns into matrices \(U_{\pm}\). The \(i\)-th row of \(U_{\pm}\) then serves as the feature vector \(y_i^{\pm}\) for node \(i\) in the corresponding view. 

After deriving the feature matrices from both views, we concatenate them horizontally to form a joint feature matrix, so that each row corresponds to the combined feature vector of a node. Our experimental results, shown in \cref{fig:f-c_diagram}, Step 4, indicate that this feature concatenation scheme outperforms the other alternatives we examined. Finally, a \(k\)-means clustering algorithm is applied to the concatenated feature matrix to determine the final clustering results, which yield the variable grouping scheme in the sub-QUBO framework.

\section{Numerical experiment results}
\label{sec:numerics}

\subsection{Dataset}
\label{subsec:dataset}
In this study, we utilize the QUBO formulation of the Max-Cut problem as a benchmark. The Max-Cut problem, an NP-hard challenge in graph theory, seeks to divide the vertices of a simple undirected graph into two distinct sets to maximize the sum of the weights of the edges that span these two sets \cite{karp2010reducibility}. This problem has significant applications in various fields, including network design \cite{goemans1995improved}, VLSI design \cite{alpert1995recent}, computational biology \cite{fortunato2010community}, combinatorial auctions \cite{andersson2000integer}, and machine learning \cite{xu2015comprehensive}. 

The Max-Cut problem can be formulated as a QUBO model and is frequently used as a benchmark for QAOA-based QUBO algorithms \cite{herrman2021impact, crooks2018performance}. In this formulation, a binary variable \(x_i\) is assigned to each vertex \(i\) in the graph, with \(x_i \in \{0,1\}\) indicating the subset to which vertex \(i\) belongs. The corresponding objective function is given by
\begin{equation}
\text{min} \quad -\sum_{(i,j) \in E} w_{ij} \, x_i \, (1 - x_j),
\end{equation}
where \(w_{ij}\) denotes the weight of the edge between vertices \(i\) and \(j\). The term \(x_i(1 - x_j)\) takes the value \(1\) precisely when \(i\) and \(j\) lie in different subsets, thereby contributing \(w_{ij}\) to the total cut weight.

Among the various graph structures studied in Max-Cut problems, regular graphs and Erdős–Rényi (ER) graphs are of particular interest. A \(k\)-regular graph has vertices all of the same degree, a property that is advantageous in network design and construction tasks involving high connectivity or minimal diameter \cite{hoory_expander_2006, tanner_recursive_1981}. An ER graph is generated by including each edge with probability \(p\) and is widely used for modeling real-world networks, such as communication or social networks \cite{cambridge_erdos_renyi}. Both graph types hold theoretical importance and serve as essential benchmarks in QAOA research \cite{farhi_quantum_2014, Larkin_Jonsson_Justice_Guerreschi_2022}.

This study considers two collections of 100-node Max-Cut problem instances. The first consists of 100 instances of 3-regular graphs, and the second comprises 100 instances of Erdős–Rényi graphs with edge probability \(p = 0.05\). Each instance is translated into a QUBO with 100 variables, a size at which classical methods such as brute force or branch and cut become intractable due to their exponential time complexity.

\subsection{Parameterization of the Algorithm}
\label{subsec:para_alg}

As shown in \cref{fig:f-c_diagram}, the sub-QUBO framework integrates a classical subroutine, a grouping method, and a quantum subroutine to address the challenges of solving QUBO problems. The classical subroutine quickly converges to local minima but may get stuck in suboptimal regions, while the quantum subroutine explores the broader solution space probabilistically. The grouping method bridges the two by partitioning variables into smaller batches for quantum processing, enhancing both optimization quality and efficiency.

For the quantum subroutine, we employ a single-layer Quantum Approximate Optimization Algorithm (QAOA), as it aligns well with the current capabilities of available quantum hardware. The single-layer QAOA circuit consists of three components: state preparation, Hamiltonian evolution of the target Ising model, and Hamiltonian evolution of the mixer Hamiltonian. The state preparation involves applying Hadamard gates to all qubits, which can be implemented efficiently. The QUBO Hamiltonian for MaxCut problems is d-sparse, where $d = \max_{v \in V} (\text{deg}(v)) + 1$, and the mixer Hamiltonian is 1-local. These structures allow for efficient Hamiltonian simulation \cite{low2017optimal, low2019hamiltonian}.

Our implementation, based on Qiskit \cite{javadi2024quantum}, simulates the quantum subroutine using 1024 shots per measurement to evaluate observables. The parameter optimization is conducted using the gradient-free COBYLA algorithm~\cite{powell1994cobyla}, chosen for its robustness in handling noisy, high-dimensional landscapes. The COBYLA optimizer is initialized with a trust region radius of $1$ at the beginning of the optimization, and gradually reduces this radius to $10^{-4}$ at termination, ensuring precise convergence to the optimal parameter set. Typically, each QAOA execution requires approximately 20 to 30 measurements to converge. Our complete simulation code is openly available under an Apache-2.0 licence at \cite{tang2025clusterqubo}.

The classical subroutine can utilize a variety of local search methods, such as greedy search or tabu search. In this study, we employ a \emph{single-bit-flip greedy descent}: starting from an initial binary vector, we scan sequentially over all variables, compute the energy change
\begin{equation}
    \Delta E_i = (1-2x_i)\Bigl(Q_{ii} + 2\sum_{j\neq i} Q_{ij}x_j\Bigr),
\end{equation}
flip \(x_i\) whenever \(\Delta E_i < 0\), and repeat the sweep until no improving move is found. The resulting routine converges deterministically to a local optimum and requires at most \(O(n^2)\) arithmetic operations per sweep. Despite its simplicity, this method has demonstrated satisfactory performance when combined with the quantum subroutine, underscoring the practical potential of this hybrid framework. A more detailed analysis of using tabu search in this context is provided in \cref{app:tabu}.

\subsection{Performance Evaluation}
\label{subsec:performance_evaluation}
\begin{figure*}
    \centering
    \includegraphics[width=17cm]{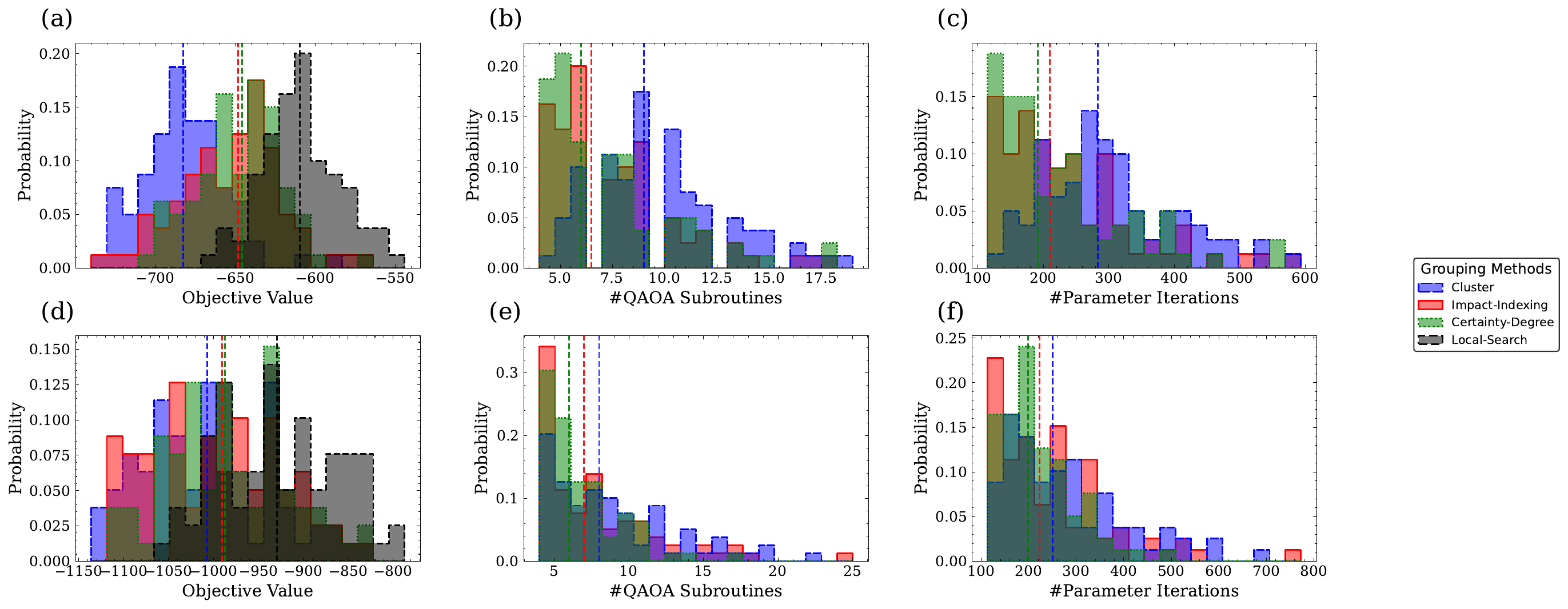}
    \caption{Performance comparison of three sub-QUBO algorithms in solving the Max-Cut problem, with a fixed sub-problem size of 24. (a, b, c) Results for 100 randomly generated Max-Cut instances based on 100-node 3-regular graphs. (d, e, f) Results for 100 randomly generated Max-Cut instances based on 100-node Erdős–Rényi (ER) graphs. (a, d) Distribution of optimal function values obtained by the three algorithms. (b, e) Distribution of the total number of QAOA subroutine calls required to solve the sub-QUBOs. (c, f) Distribution of the total number of parameter iterations across all QAOA instances (each QAOA instance involves multiple iterations to optimize its ansatz parameters until convergence).}
    \label{fig:panel_fixed_d}
\end{figure*}

We compared three hybrid QUBO solvers with different grouping methods—Impact-Indexing, Certainty-Degree, and our proposed clustering—in solving Max-Cut problems generated from 100-node 3-regular and Erdős–Rényi graphs. Apart from their distinct grouping approaches, all three solvers share the same components: they employ QAOA with depth \(p=1\) to tackle each sub-QUBO (of fixed size \(d=24\)), followed by a local greedy search as the classical subroutine. The results, shown in \cref{fig:panel_fixed_d}, reveal that the clustering grouping method consistently reaches a lower objective value than the other two methods, while incurring a comparable number of quantum subroutine calls to the Impact-Indexing method. These observations hold for both the 3-regular and Erdős–Rényi graph cases.

\begin{figure*}
    \centering
    \includegraphics[width=17cm]{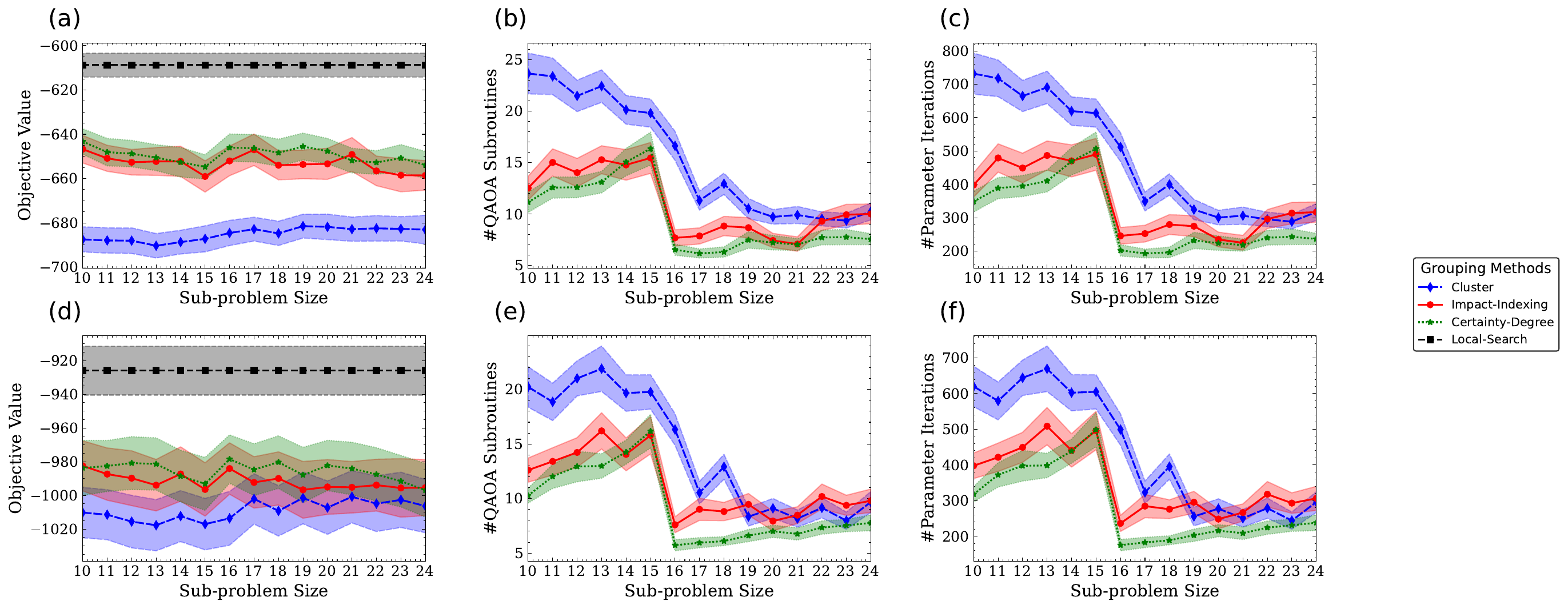}
    \caption{Performance comparison and computational resource requirements for three sub-QUBO algorithms solving Max-Cut problems with sub-problem sizes ranging from 10 to 24 qubits. The algorithms use Impact-Indexing (red), Certainty-Degree (green), and clustering (blue) as grouping methods. (a,d) Minimum objective function values for 100 randomly generated Max-Cut problems based on (a) 100-node 3-regular graphs and (d) 100-node Erdős–Rényi graphs. Lower values indicate better performance. (b,e) Number of quantum QAOA subroutine calls versus sub-problem size for (b) 3-regular graphs and (e) Erdős–Rényi graphs. (c,f) Total number of QAOA iterations versus sub-problem size for (c) 3-regular graphs and (f) Erdős–Rényi graphs. Shaded areas represent 95\% confidence intervals.}
    \label{fig:panel_scale}
\end{figure*}

For the same dataset of randomly generated 3-regular weighted graphs and ER graphs, we further investigate the performance versus the sub-QUBO size $d$ ranging from 10 to 24. The results of the experiment are illustrated in \cref{fig:panel_scale}. As the sub-QUBO size increases from 10 to 24, the minimum objective function values obtained by both hybrid algorithms remain almost unchanged. However, the number of QAOA subroutine calls decreases significantly. 
This observation aligns with the intuition that larger sub-QUBO sizes enable the QAOA subroutine to explore more of the solution space per iteration, thereby reducing the need for multiple iterations and classical post-processing. As the sub-QUBO size increases, each QAOA subroutine becomes more effective, leading to fewer required iterations. In the extreme case, when the available qubit count matches or exceeds the full QUBO problem size, a single QAOA subroutine call suffices to solve the entire problem without decomposition.

Our numerical experiments reveal that for smaller sub-QUBO sizes, the clustering method requires more QAOA subroutine calls compared to other methods. However, this difference decreases as the sub-QUBO size increases, eventually becoming negligible compared to statistical fluctuations. This suggests that for sufficiently large sub-QUBO sizes solvable by a powerful quantum computer, the clustering method may require an equal or even lower number of QAOA subroutine calls compared with other grouping methods within the sub-QUBO formalism.

While the mean value of the objective function follows the same relative trend, the solutions for Max-Cut problems on ER graphs exhibit larger fluctuations compared to those for 3-regular graphs. These fluctuations are illustrated by the wider distribution of optimal function values in \cref{fig:panel_fixed_d}(d) compared to \cref{fig:panel_fixed_d}(a), as well as the larger confidence intervals shown in \cref{fig:panel_scale}(d) compared to \cref{fig:panel_scale}(a). This increased variability is likely due to the higher randomness in the generation of ER graphs compared to the more structured 3-regular graphs.

In \cref{fig:panel_scale}(b) and (e), we observe a sharp decrease in the number of quantum subroutine calls when the sub-QUBO size increases from $15$ to $16$ for the Certainty-Degree and Impact-Indexing grouping methods. Further numerical experiments on varying Max-Cut problem sizes using three grouping methods reveal that both Certainty-Degree and Impact-Indexing methods exhibit sudden drops in the number of subroutine calls at specific ratios of sub-QUBO size to problem size. In contrast, the clustering method demonstrates a more consistent behavior, with the number of quantum subroutine calls approximately inversely proportional to the sub-QUBO size across all problem sizes. Detailed discussions of these results are conducted in \cref{app:calls}.

\section{Conclusion}
\label{sec:conclusion}

In this study, we enhance the sub-QUBO formalism by introducing clustering as the grouping method. We propose a measure for variable correlation in QUBO problems and utilize multi-view spectral clustering to group variables into sub-QUBOs. Our numerical simulations on Max-Cut problems, using randomly generated 3-regular graphs and Erdős–Rényi graphs, demonstrate that clustering outperforms existing grouping methods in obtaining optimal objective function values. The results indicate that while the size of sub-QUBOs does not significantly impact solution quality, smaller sub-QUBOs generally require more quantum subroutine calls for convergence. Despite yielding superior objective function values, the clustering method maintains efficiency, requiring a comparable number of quantum subroutine calls to existing methods.
These findings are consistent across various classical subroutines, including greedy local search and tabu search. Future work could explore the application of this method to other combinatorial optimization problems and further refine the clustering algorithms to enhance performance.

% \bibliographystyle{unsrt}
% \bibliography{bib/sample.bib}

\section*{Acknowledgements}
This work was supported by Beijing Municipal Science \& Technology Commission, Beijing Science and Technology Plan ($Z241100004224034$).

\appendix

\section{Tabu Search}
\label{app:tabu}
The Tabu Search algorithm\cite{glover1998tabu,glover_diversification-driven_2010}, as described in \cref{alg:tabu_search}, is an optimization technique aimed at minimizing the QUBO objective function.  
The algorithm seeks to find the best solution by iteratively improving the current solution while avoiding cycles and revisits to previously encountered solutions through the use of a tabu list.

\begin{algorithm}[H]
\caption{Tabu Search for QUBO Minimization}
\label{alg:tabu_search}
\begin{algorithmic}[1]
\State \textbf{Input:} Initial solution $\mathbf{s}$, QUBO matrix $Q$, max iterations $iter\_max$, tabu tenure $nTabu$, target energy $target$
\State \textbf{Output:} Best solution found $\mathbf{s}^*$ and its energy $E^*$
\State Initialize $\mathbf{s}^*$ as $\mathbf{s}$ and $E^*$ as $E(\mathbf{s})$ \label{line:init_solution}
\State Initialize tabu list $TabuK$ with zeros \label{line:init_tabu_list}
\State Initialize bit flip costs $flip\_cost$ and sort index $index$ \label{line:init_flip_cost}
\State Set $Vlastchange$ to $E^*$ and $bit\_flips$ to 0 \label{line:init_vars}

\While{$bit\_flips < iter\_max$} \label{line:main_loop}
    \State Set $neighbour\_best$ to infinity and $improved$ to false
    \For{each bit $k$ in $index$}
        \If{bit $k$ is not in tabu list} \label{line:check_tabu}
            \State Calculate new energy with bit $k$ flipped \label{line:calc_energy}
            \If{new energy is better than $E^*$}
                \State Update solution, energy, and tabu list \label{line:update_best}
                \State Set $improved$ to true and break loop
            \ElsIf{new energy is the best among neighbors} \label{line:check_best_neighbor}
                \State Update best bit to flip \label{line:update_best_bit}
            \EndIf
        \EndIf
    \EndFor
    
    \If{no improvement found} \label{line:no_improvement}
        \State Flip the best bit found and update energy and tabu list \label{line:flip_best_bit}
    \EndIf
    
    \State Decrement all tabu tenures \label{line:update_tabu_list}
    \State Increment $bit\_flips$ \label{line:increment_bit_flips}
    
    \If{best energy $E^*$ is less than or equal to target} \label{line:check_target}
        \State Exit loop \label{line:exit_loop}
    \EndIf
\EndWhile

\State \Return best solution $\mathbf{s}^*$ and its energy $E^*$ \label{line:return_solution}
\end{algorithmic}
\end{algorithm}

The tabu search algorithm maintains a tabu list to prevent recently flipped bits from being immediately flipped back, thus avoiding cycles. The algorithm initializes with a starting solution and iteratively improves it while respecting the tabu constraints.

The main loop of the algorithm (line \ref{line:main_loop}) continues until either a maximum number of iterations is reached or a target energy is achieved. In each iteration, the algorithm attempts to flip bits that are not in the tabu list. If a bit flip improves the best-known solution, it is accepted and marked as tabu for a certain number of iterations (line \ref{line:update_best}). If no improvement is found, the algorithm selects the best non-tabu move available. The tabu status of bits is updated after each iteration.

This process allows the algorithm to explore the solution space effectively, escaping local optima by occasionally accepting moves that do not immediately improve the solution. The search terminates when either the iteration limit is reached or the target energy is achieved.

\begin{figure}[H]
    \centering
    \includegraphics[width=\linewidth]{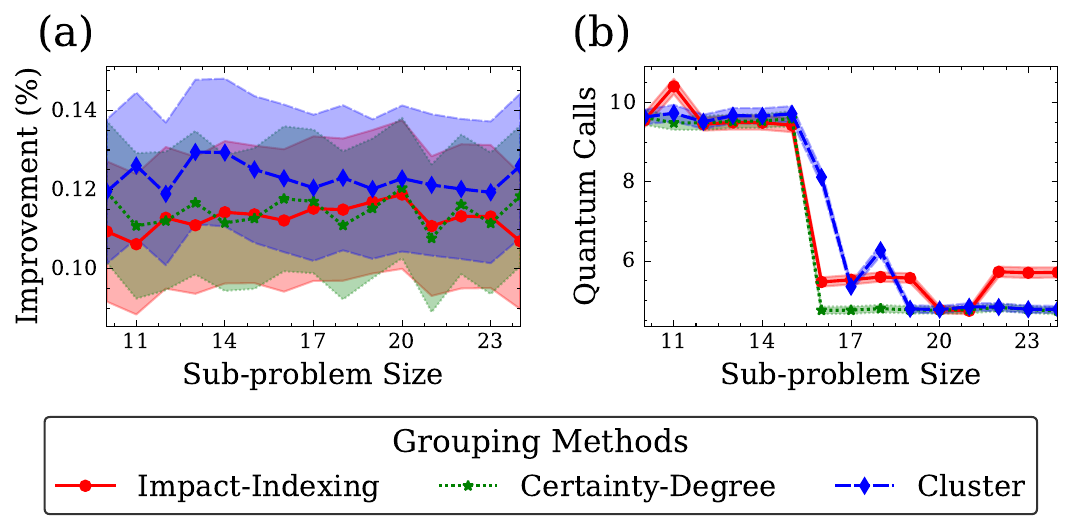}
    \caption{Performance dependence on sub-QUBO size $d$ in Max-Cut problems for 100-node 3-regular graphs, comparing clustering and Impact-Indexing as grouping methods. (a) Average minimum objective function value found by both algorithms. (b) Average number of QAOA subroutine calls required by each algorithm. Both subfigures include a 95\% confidence interval. }
    \label{fig:tabu}
\end{figure}

The results presented in \cref{fig:tabu} demonstrate that the qualitative findings obtained using tabu search as the classical subroutine are consistent with those observed when using a greedy local search. However, it is important to note that tabu search, being a more sophisticated optimization technique, generally achieves better results on its own compared to simpler methods like greedy local search. Consequently, while the relative performance of different grouping methods (such as clustering slightly outperforming Impact-Indexing) remains consistent, the overall improvement contributed by the quantum subroutine appears less pronounced when tabu search is employed as the classical component. 

\section{Algorithm Complexity Analysis}
\label{app:calls}

For a QUBO or Max-Cut instance on an $n$-vertex graph with $m$ edges, a $p$-layer QAOA circuit applies one entangling gate per edge and one single-qubit rotation per qubit in each layer, leading to a total gate count of $O(pm)$ and a circuit depth of $O(p\Delta)$ on hardware with maximum degree $\Delta$~\cite{farhi_quantum_2014,MDPI2021Depth}. On fully connected quantum devices, where entangling gates can be applied in parallel, this depth reduces to $O(p)$~\cite{Hastings2019Graph}.  

The cost of evaluating a single parameter vector to additive precision $\varepsilon$ requires $O(1/\varepsilon^2)$ measurements due to Hoeffding’s bound~\cite{Hoeffding1963}. Given that the outer-loop optimizer requires $N_{\mathrm{eval}}$ such evaluations, where heuristic methods typically keep $N_{\mathrm{eval}}$ polynomial for fixed $p$~\cite{Zhou2020PRX,Hardware2022Scaling}, the total runtime depends critically on the number of these evaluations. Additionally, in the large system limit, analytical expressions can be evaluated in $O(16^p)$ time, reflecting the exponential scaling with $p$~\cite{Goldstone2022SK}.  

With a fixed $p$, the quantum subroutine's time complexity remains effectively constant, and the primary factor influencing overall runtime becomes the number of quantum subroutine calls. The following section examines how this call count scales with the problem size $N$ and the sub-QUBO size $d$ under different grouping strategies, offering insights into the practical scalability of the approach.

As illustrated in \cref{fig:panel_scale} and \cref{fig:tabu}, the number of quantum subroutine calls decreases significantly when the sub-QUBO size increases from 15 to 16 for the Impact-Indexing and Certainty-Degree grouping methods. In contrast, the clustering method introduced in this work exhibits a smoother relationship between sub-QUBO size and the number of quantum subroutine calls. To further investigate this trend, we conduct experiments on Max-Cut problems for 3-regular graphs with sizes ranging from 80 to 180 nodes. The three grouping methods were employed, and the number of quantum subroutine calls was recorded as the sub-QUBO size varied from 5 to 24.
\begin{figure}
    \centering
    \includegraphics[width=\linewidth]{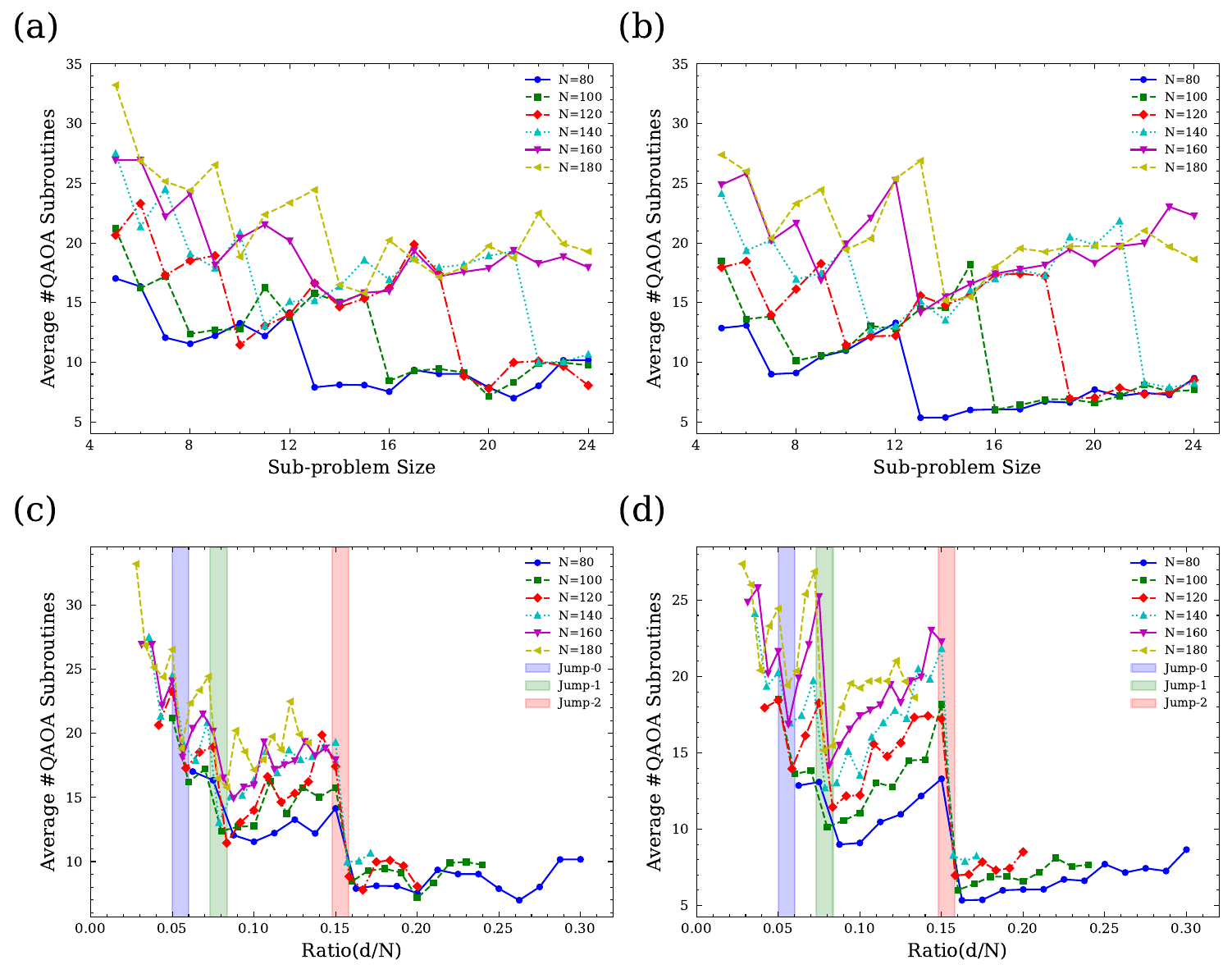}
    \caption{Algorithm complexity performance for Impact-Indexing and Certainty-Degree grouping methods in solving Max-Cut problems. (a,b) Variation of average quantum subroutine calls with different sub-QUBO sizes for (a) Impact-Indexing and (b) Certainty-Degree methods. (c,d) Relationship between average quantum subroutine calls and the size ratio, $\frac{d}{N}$, for (c) Impact-Indexing and (d) Certainty-Degree algorithms. The curves exhibit changes at three distinct values. The intervals containing these critical points are shaded in different colors, denoted as $critical-{1,2,3}$ respectively.}
    \label{fig:scaling_jump}
\end{figure}

\cref{fig:scaling_jump} demonstrates that under Impact-Indexing and Certainty-Degree grouping methods, the number of quantum subroutine calls as a function of sub-QUBO ratio $\frac{d}{N}$ displays sudden drops at certain values. These critical points are shaded in \cref{fig:scaling_jump} and detailed in \cref{tab:critical_points}.

\begin{table}[H]
\centering
\begin{tabular}{cccc}
\toprule

\textbf{$N$ } & \textbf{ $d_{critical}^{(1)}$ (Ratio) } & \textbf{ $d_{critical}^{(2)}$ (Ratio)} & \textbf{ $d_{critical}^{(3)}$ (Ratio) } \\

\midrule
80  & {} & \textbf{7} \(\scriptstyle (0.0875)\) & \textbf{13} \(\scriptstyle (0.1625)\) \\
100 & \textbf{6} \(\scriptstyle (0.06)\) & \textbf{8} \(\scriptstyle (0.08)\) & \textbf{16} \(\scriptstyle (0.16)\) \\
120 & \textbf{7} \(\scriptstyle (0.0583)\) & \textbf{10} \(\scriptstyle (0.0833)\) & \textbf{19} \(\scriptstyle (0.1583)\) \\
140 & \textbf{8} \(\scriptstyle (0.0571)\) & \textbf{11} \(\scriptstyle (0.0786)\) & \textbf{22} \(\scriptstyle (0.1571)\) \\
160 & \textbf{9} \(\scriptstyle (0.0563)\) & \textbf{13} \(\scriptstyle (0.0813)\) & {} \\
180 & \textbf{10} \(\scriptstyle (0.0556)\) & \textbf{14} \(\scriptstyle (0.0778)\) & {} \\

\bottomrule
\end{tabular}
\caption{Critical points for Impact-Indexing and Certainty-Degree grouping methods. $N$ is the problem size (80-180 nodes). Values show critical sub-QUBO sizes $d$, with ratios \(\frac{d}{N}\) in parentheses.}
\label{tab:critical_points}
\end{table}

\begin{figure*}
    \centering
    \includegraphics[width=17cm]{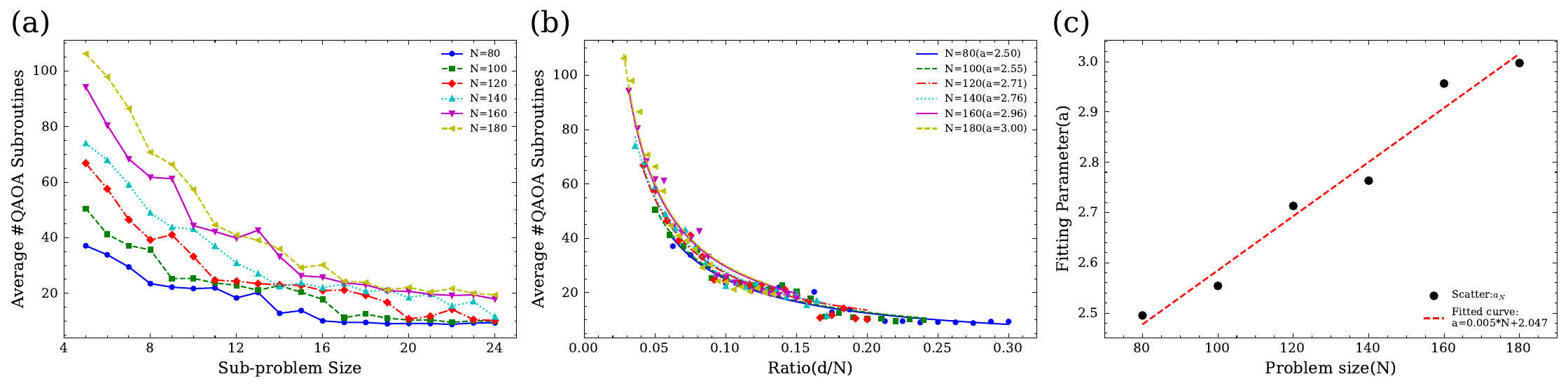}
    \caption{Complexity analysis of the clustering-based Sub-QUBO algorithm for solving Max-Cut problems. Experiments involve 80 randomly generated Max-Cut problems based on 3-regular graphs with different problem sizes, $N$, equivalent to the number of nodes in the graph. The size of each sub-problem is denoted as $d$, ranging from $5$ to $24$. The ratio is defined as $d/N$. (a) Relationship between average number of quantum subroutine calls and sub-QUBO size. Six curves represent results for different problem sizes. (b) Average quantum subroutine calls versus ratio for different problem sizes. Scatter plots show results for $N\in\{80,100,120,140,160,180\}$. Curves are fitted using the function: quantum subroutine calls $=\frac{a}{ratio}= \frac{a}{d/N}$, where $a$ is the fitting parameter. (c) Relationship between fitting parameter $a$ and problem size $N$. Scatter plot shows experimental results, and the line represents a linear fit.}
    \label{fig:scaling_smooth}
\end{figure*}

\cref{fig:scaling_smooth} illustrates that under the cluster grouping method, for a fixed total problem size, the number of quantum subroutine calls and the proportion of the sub-QUBO size within the total problem size follow an inverse proportional function. Their product increases linearly with the total problem size. 

Consequently, the number of quantum subroutine calls can be approximated as $\frac{0.005N^2 + 2.047N}{d}$, where $N$ is the problem size and $d$ is the sub-QUBO size. The number of calls follows $O\left(\frac{N^2}{d}\right)$. Comparing this with existing grouping methods, we conclude that the cluster method requires a comparable number of quantum subroutine calls. However, the decay in the number of calls with respect to the $\frac{d}{N}$ ratio is more gradual for the cluster method.

\end{document}